\documentclass[traditabstract]{aa} 

\usepackage[colorlinks=true, linkcolor=blue, citecolor=blue, urlcolor=blue]{hyperref}
\usepackage{graphicx}
\usepackage{txfonts}
\usepackage{natbib}

\begin{document}

   \title{High resolution X-ray spectra of the compact binary supersoft X-ray source CAL 87}
 \author{Songpeng Pei\inst{1}
          \and
          Xiaowan Zhang\inst{1}
          \and
          Qiang Li\inst{2, 3}
          \and
          Ziwei Ou\inst{4}
          }

   \institute{School of Physics and Electrical Engineering, Liupanshui Normal University, Liupanshui, Guizhou, 553004, China\\
         \email{songpengpei@outlook.com}
         \and
             Qiannan Normal University for Nationalities, Duyun 558000, China
         \and
             Qiannan Key Laboratory of Radio Astronomy, Guizhou Province, Duyun 558000, China
         \and
             Tsung-Dao Lee Institute, Shanghai Jiao Tong University, Shanghai 201210, China
             }
   \date{Received 00.00.2024; accepted 00.00.2024}
   
\abstract{In this study, we present an analysis of the archival X-ray data of the eclipsing supersoft X-ray binary CAL 87 observed with the {\it Chandra} Advanced CCD Imaging Spectrometer (ACIS) camera and Low Energy Transmission Grating (LETG) in 2001 August and with {\it XMM-Newton} in 2003 April. The high resolution X-ray spectra are almost unchanged in the two different dates. The average unabsorbed X-ray luminosity during the exposure was 4.64$-$5.46$\times10^{36}$ ergs s$^{-1}$ in 2001 and 4.54$-$4.82 $\times10^{36}$ ergs s$^{-1}$ in 2003, with prominent and red-shifted emission lines, mostly of nitrogen, oxygen, iron and argon, contributing to at least 30\% of the X-ray flux. The continuum X-ray flux is at least an order of magnitude too small for a hot, hydrogen burning WD. However, the continuum flux is consistent with Thomson-scattering reflecting about 5\% of the light of a hydrogen burning WD with effective temperature of 800,000 K, and a mass of $\sim$ 1.2 M$_\odot$. It has been noted before that a large Thomson-scattering corona explains the X-ray eclipse of CAL 87, in which the eclipsed region is found to be of size the order of a solar radius. The emission lines originate in an even more extended region, beyond the eclipsed central X-ray source; the emission spectrum is very complex, with unusual line ratios.}


%
\titlerunning{High resolution X-ray spectra of CAL 87}
\maketitle

\section{Introduction} \label{sec:intro}

CAL 87 is an eclipsing compact binary supersoft X-ray source (CBSS) in the Large Magellanic Cloud (LMC) that was discovered with the imaging proportional counter (IPC) of the Einstein Observatory \citep{1981ApJ...248..925L}. The orbital period of CAL 87 is 10.6 hours \citep{1989MNRAS.241P..37C, 1990ApJ...350..288C, 1993PASP..105..863S}. CBSS are characterized by black body-like spectra with temperatures $\sim 20-100$ eV and X-ray luminosities $\sim 10^{35}-10^{38}\, {\rm erg\,s^{-1}}$ \citep{1996LNP...472.....G}. \citet{1992A&A...262...97V} suggested that the X-ray emission of CBSS is not transient or recurrent and cannot be associated with nova events. The luminous supersoft X-ray emission stems from stable nuclear burning on the surface of WDs, which requires high mass transfer ($\gtrsim 10^{-7}$ M$_\odot$ y$r^{-1}$) from the donor star in an interacting binary. In this model mass transfer is not permanently stable, and the burning is a recurrent phase in the evolution of a binary with a mass donor of larger mass than the WD. Mass transfer eventually becomes dynamically unstable and there are cycles during which the WD's envelope increases its radius to a red giant configuration.

Using the UV spectra, \citet{1995AJ....110.2394H} estimated that the disk bolometric luminosity is $\sim$ $10^{36}$ ergs s$^{-1}$. \citet{1998ApJ...503L.143A} and \citet{2001ApJ...550.1007E} interpreted the luminosity of the system as due to the atmosphere of a WD undergoing stable hydrogen burning in shell, and tried to fit the broad band X-ray spectra observed by the ASCA Solid-State Spectrometer \citep[SIS;][]{1994PASJ...46L..37T}, which has an
energy resolution of $\Delta E/E$ = $\sim$ 10\% at 0.5 keV, as such. Because of an eclipse in the X-rays, \citet{2001ApJ...550.1007E} were able to find out that the eclipsed region has to be of the order of magnitude of a solar radius in size, and concluded that what is detected in X-rays is Thomson scattered radiation in a corona surrounding a WD as primary star, while the WD itself is never observed.

However, observations done with the {\it Chandra} ACIS camera \citep{2003SPIE.4851...28G} and the LETG \citep{2008HEAD...10.0403D} in 2001, and later with the Reflection Grating Spectrometers \citep[RGS;][]{2001A&A...365L...7D} instrument of the European Space Agency’s X-ray Multi-Mirror Mission ({\it XMM-Newton}) observatory \citep{2001A&A...365L...1J} in 2003, surprisingly failed to detect a WD spectrum and indicated that the X-ray flux is in large part due to emission lines, that were not resolved with the CCD-type instruments used earlier. \citet{2004RMxAC..20...18G} suggested that the LETG spectra are due to a recombination component and a resonant scattering component, and are formed by a hidden ionizing central source that they inferred to have an absolute unabsorbed luminosity $\sim$ $5 \times10^{37}$ ergs s$^{-1}$ and an effective temperature in the $\sim$ $50-100$ eV range. \citet{2004RMxAC..20..210O} identified the emission lines in the spectra of {\it XMM-Newton} RGS data, and based on the red shift ($700-1200$ km s$^{-1}$) of the lines, they suggested that these lines originated from a wind. However, these articles were two short conference presentations and no quantitative models of the grating spectra followed. \citet{2013A&A...559A..50N} classified Cal 87 and U Sco as a subclass of SSS called the SSe class. In this class, the X-ray spectrum features a weak blackbody-like continuum without absorption lines, and the emission lines are at least as strong as the continuum. Moreover, the emission lines in the SSe class are strongest where the continuum is most intense. \citet{2024ApJ...960...46T} fitted the spectra of {\it XMM-Newton} data by using a corona model based on the radiative transfer of the corona surrounding the WD atmosphere plus a collisionally ionized plasma model, while the energy ranges of 0.565$-$0.569 keV (21.790$-$21.944 \AA) and 0.715$-$0.740 keV (16.755$-$17.341 \AA) was ignored in their final fitting.

\citet{2004RMxAC..20..210O} noticed that the eclipse is energy dependent, and is deeper at lower energy. Using the {\it XMM-Newton} data, \citet{2014ApJ...792...20R} studied the X-ray eclipse geometry of CAL 87 in the {\it XMM-Newton} data. They found that the continuum emission dominates the decrease of flux during eclipses, and that the emission lines are not eclipsed and originate in a more extended region. While an eclipse model is also shown in \citet{2001ApJ...550.1007E}, these authors assumed that the Thomson scattering corona is eclipsed
by the secondary blocking the supersoft X-ray light. \citet{2014ApJ...792...20R} suggested
 instead that an accretion disk with an asymmetric feature eclipses the source of the continuum supersoft radiation. \citet{2013A&A...559A..50N} also suggested that when viewed edge-on, the accretion disc blocks the central hot source, however, Thomson scattering allows the continuum emission to remain visible even during the total obscuration of the central hot source. The optical eclipse, shifted in phase by 3\% \citep{2004RMxAC..20..210O} was explained by \citet{1997A&A...318...73S} as due to a secondary star more massive than the compact object, eclipsing the accretion disk, which also has an asymmetric hot spot where the mass inflow impinges the disk, producing a non-symmetric light curve in ingress and regress.

CAL 87 is an eclipsing source in optical
 and X-rays, with an inclination larger than $70^{\circ}$, 
 but measuring the mass of the binary components 
has not been possible so far. Recent papers tend to disregard \citet{1992A&A...262...97V} arguments in favor of recurrent stable mass transfer with a more massive secondary than the compact object (a G to F star of $\sim$ 1.5 M$_\odot$ mass), as well as the possibility of a different compact object than a WD. \citet{2014ApJ...792...20R} favor a low mass secondary (of less than 1 M$_\odot$ and of lower mass than the WD) and find evidence that the most likely values of the inclination $i$ and mass ratio are $i$ = $82^{\circ}$ and $q$ = 0.25.

In this work, we reanalyzed previously published data obtained with {\it Chandra} \citep{2004RMxAC..20...18G} and {\it XMM-Newton} \citep{2004RMxAC..20..210O, 2013A&A...559A..50N, 2014ApJ...792...20R, 2024ApJ...960...46T}. The X-ray spectra of CAL 87 are complex, and the obtained high resolution X-ray spectra of CAL 87 are precious and with features that could be used to sensibly constrain models. However, previous attempts at spectral modeling for these high-resolution X-ray spectra have not been successful. We aim to explore the high-resolution X-ray spectra of CAL 87 to identify models that can adequately fit the data. Our findings indicate that an atmospheric model combined with a collisional plasma model can fit the blackbody-like continuum and some of the emission lines in the X-ray spectra of CAL 87. We present a brief introduction of CAL 87 in Section \ref{sec:intro}. We describe the details of the observations and briefly recall the data reduction process in Section \ref{sec:observation}. We present a detailed analysis of the grating spectra in Section \ref{sec:spectral}. Our findings are discussed in Section \ref{sec:discussion}, and we summarize our conclusions in \ref{sec:conclusions}.

\section{Observation and data reduction}
\label{sec:observation}
CAL 87 was observed with the ACIS and LETG instruments of {\it Chandra} on 2001 August 13 $-$ 14, for nearly 100 ksec, covering two full orbital cycles and three eclipses (PI: Greiner; ID: 1896). We extracted the {\it Chandra} 1st order spectrum and zero order light curve with {\it Chandra} Interactive Analysis of Observation 
\citep[CIAO;][]{2006SPIE.6270E..1VF} v4.12 \footnote{https://cxc.cfa.harvard.edu/ciao/download/} using the calibration files CALDB v4.9.0, and all the standard procedures of data reprocessing and reduction in CIAO science threads were used.

\begin{figure} 
\includegraphics[width=88mm]{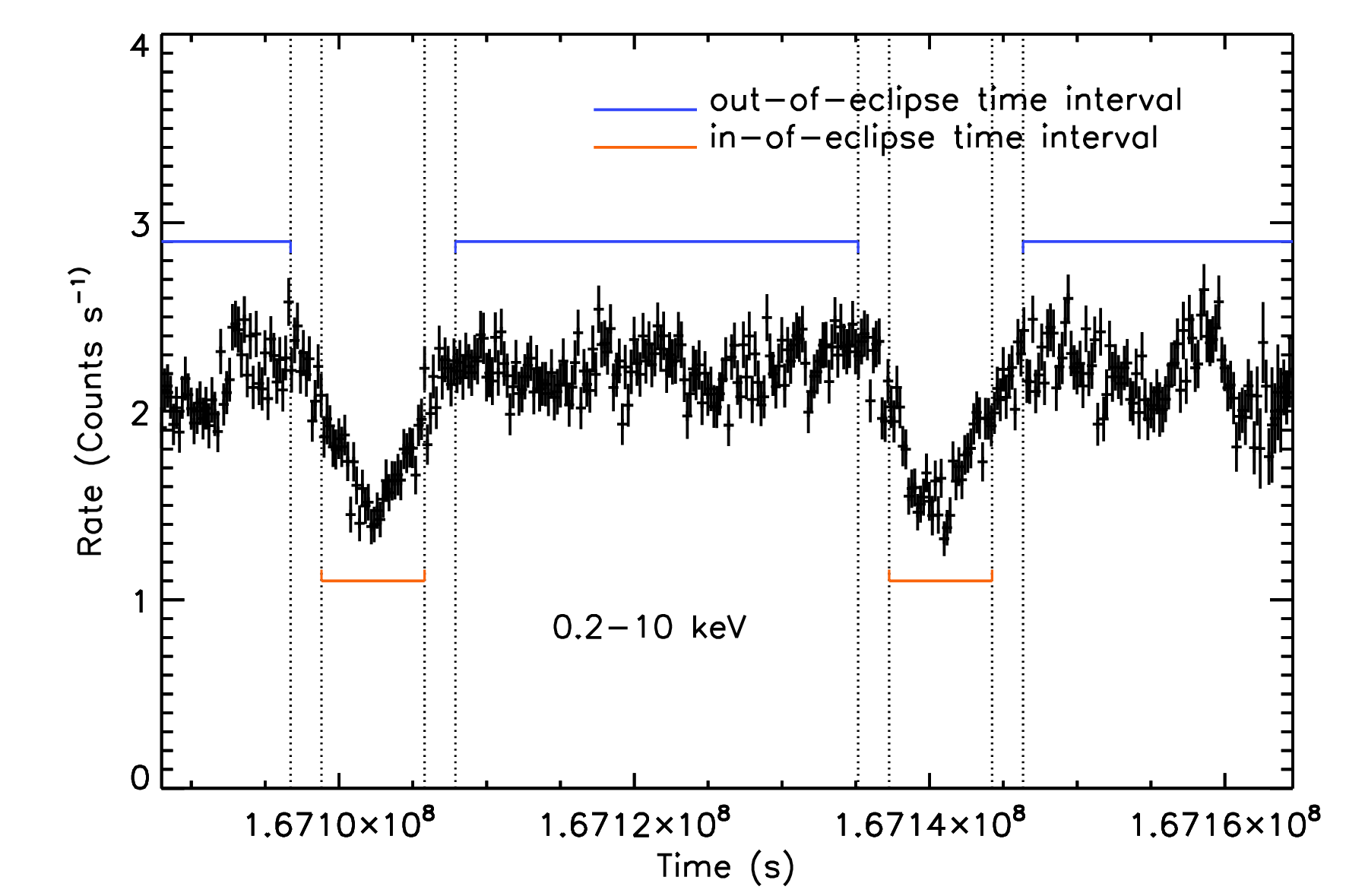}
\caption[The light curve of the {\it XMM-Newton} EPIC-pn observation]{The light curve of the {\it XMM-Newton} EPIC-pn observation of CAL 87 in the 0.2 $-$ 10 keV, binned with 200 s per bin. The blue lines show the time intervals in which
we extracted the out of eclipses spectrum, and the blue lines show the time intervals in which we extracted the in of eclipse
spectrum. For the comparison, we neglected 
 the spectrum observed in a 2100 s interval at the beginning and end of the eclipses.}
\label{fig:lc_cal87}
\end{figure}

CAL 87 was also observed with all the {\it XMM-Newton} instruments: the RGS1, the RGS2, the European Photon Imaging Camera-pn \citep[EPIC-pn;][]{2001A&A...365L..18S}, the EPIC-Metal Oxide Semi-conductor 1 (MOS1), the EPIC-MOS2 \citep{2001A&A...365L..27T} and the Optical Monitor \citep[OM;][]{2001A&A...365L..36M} on 2003 April 18 $-$ 19 for nearly 80 ksec, covering two orbital cycles and two eclipses (PI: Ebisawa; ID: 0153250101), as shown in Fig.~\ref{fig:lc_cal87}. The {\it XMM-Newton} data reduction was performed using the Science Analysis System (SAS) version 20.0.0 package and Current Calibration Files (CCF). We extracted the RGS1 and RGS2 gratings spectra using the updated calibration files, and all the standard procedures of data reprocessing and reduction in SAS Data Analysis Threads. We co-added and averaged the two background subtracted grating spectra with the SAS task RGSCOMBINE.

We used XSPEC v. 12.12.1 \citep{1996ASPC..101...17A, 2003HEAD....7.2210D} to fit the spectra. For the regions below 13 \AA\ and above 34 \AA\ the signal-to-noise ratio of the background subtracted spectra is very low, so we used the 13-34 \AA\ range to perform the spectral analysis.

\begin{figure*}
\includegraphics[width=92mm]{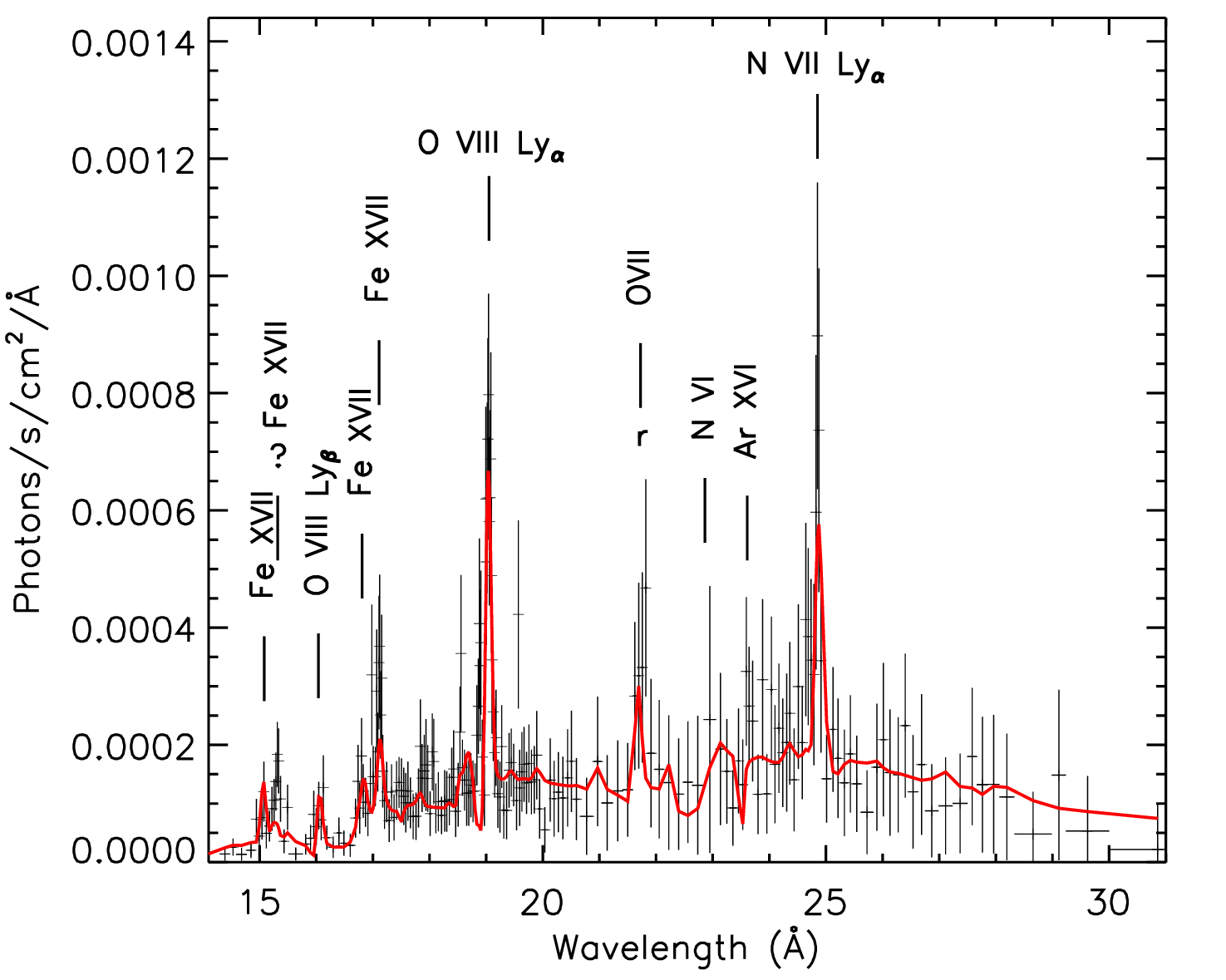}
\hspace{0.01cm}
\includegraphics[width=92mm]{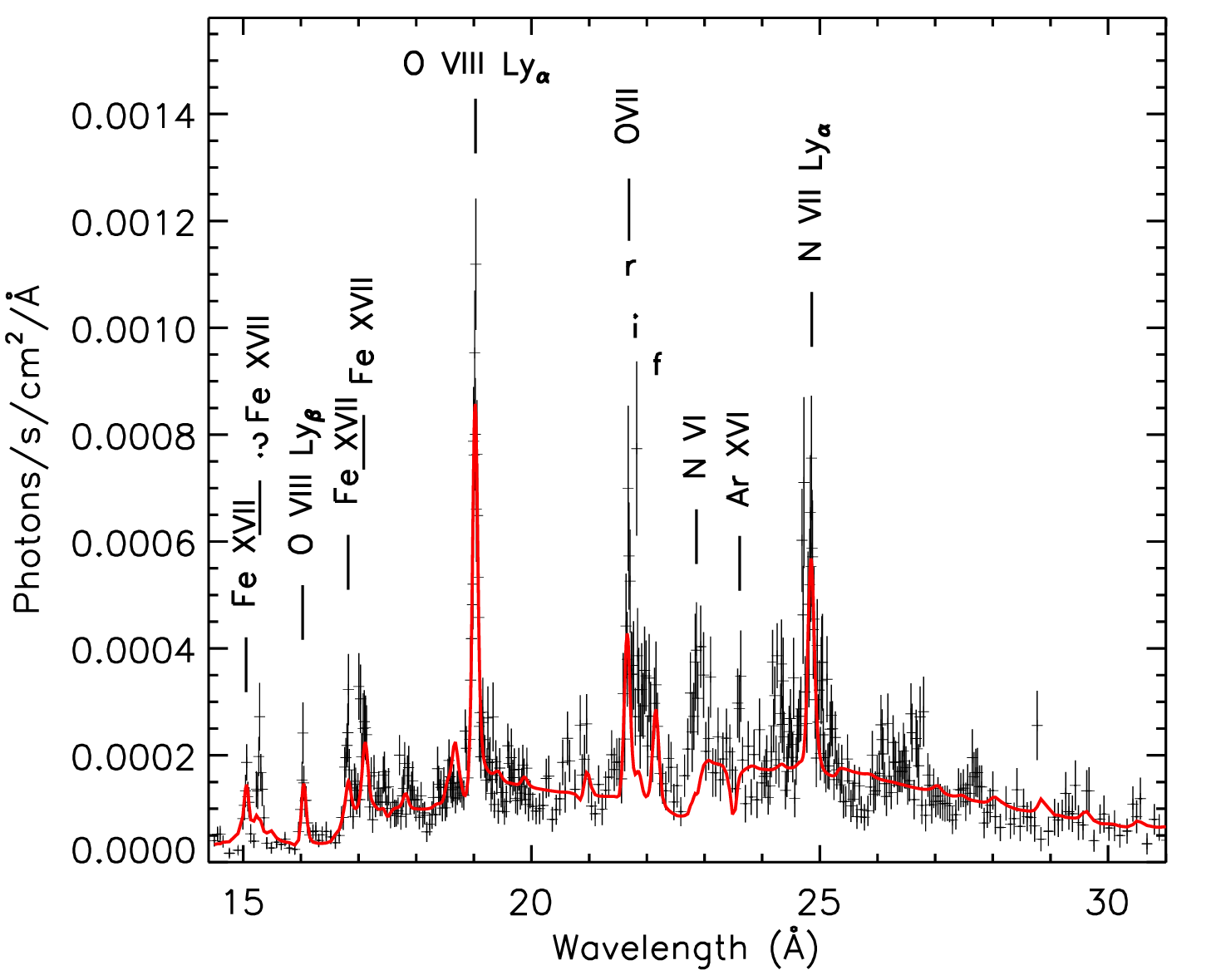}
 \caption[The {\it Chandra} spectra and {\it XMM-Newton} spectra]{The {\it Chandra} LETG spectrum (left) and the averaged {\it XMM-Newton} RGS spectrum (right) of CAL 87, fitted with an atmosphere model and a collisional ionization plasma model, whose characteristics and parameters are given in Table~\ref{table:parameters bvapec cal87}. The fit is shown with the red solid line.}
\label{fig:bvapec}
\end{figure*}

\begin{figure*}
\includegraphics[width=92mm]{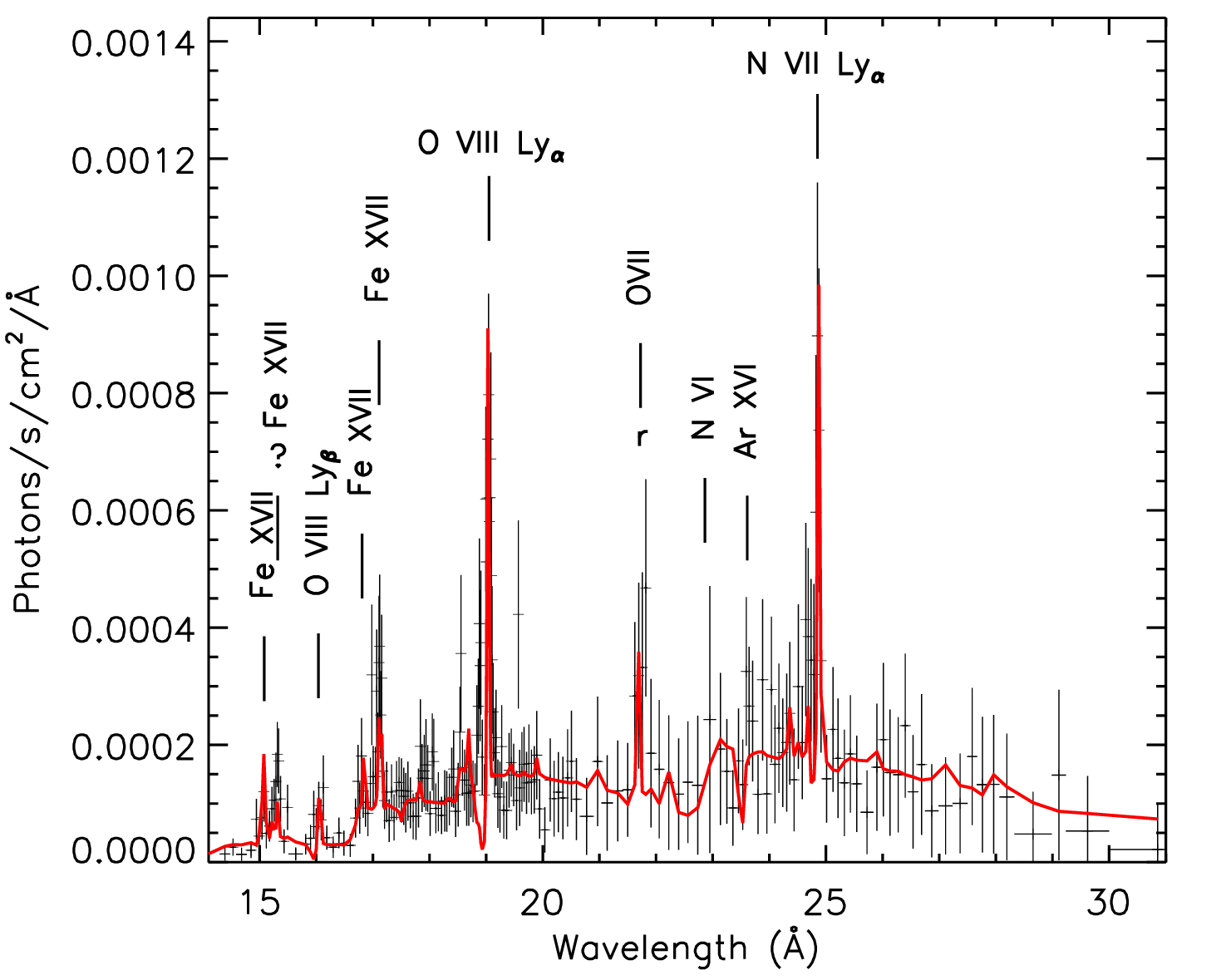}
\hspace{0.01cm}
\includegraphics[width=92mm]{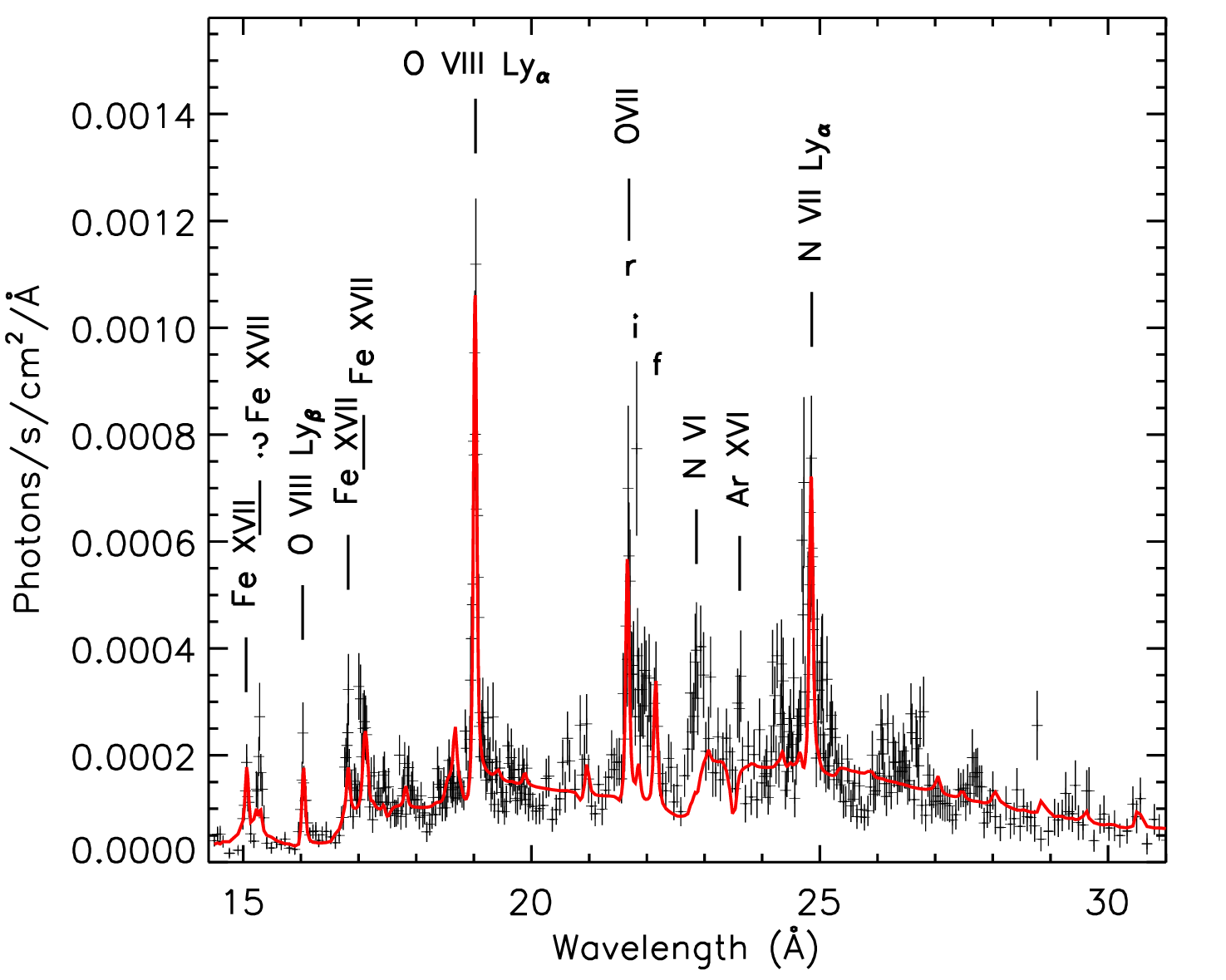}
 \caption[The {\it Chandra} spectra and {\it XMM-Newton} spectra]{The {\it Chandra} LETG spectrum (left) and averaged {\it XMM-Newton} RGS spectrum (right) of CAL 87, fitted by using an atmosphere model and a non-equilibrium ionization collisional plasma model, whose characteristics and parameters are given in Table~\ref{table:parameters bvapec cal87}. The fit is shown with the red solid line.}\label{fig:vnei}
\end{figure*}

\begin{figure} 
\includegraphics[width=0.68\columnwidth,angle=270,clip]{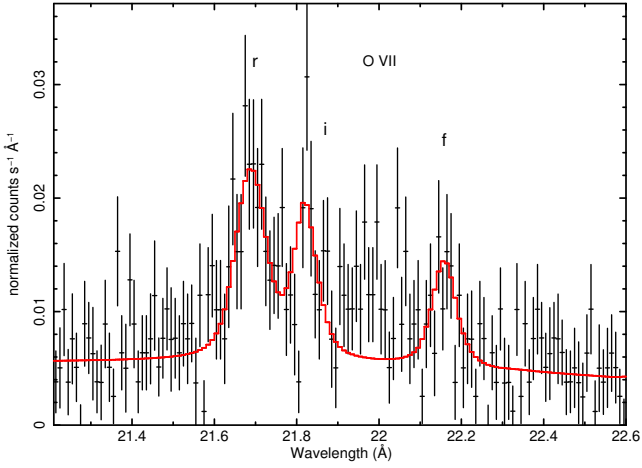}
\caption[O VII He-like triplet]{The O VII He-like triplet, unbinned, as measured in the averages RGS1 and RGS2 spectra of CAL 87, (black data points), and a fit with Gaussians for the lines, and a power law for the continuum (red solid line).}
\label{fig:triplet}
\end{figure}

\begin{figure*}
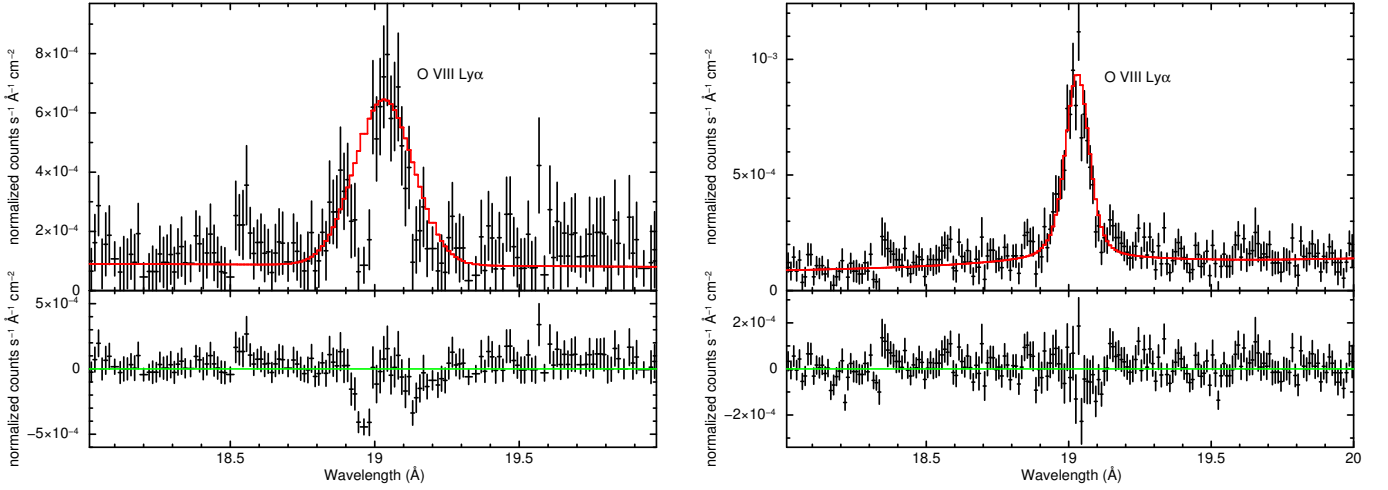

\includegraphics[width=64mm,angle=270,clip]{line_6_ch.eps}
\hspace{0.01cm}
\includegraphics[width=64mm,angle=270,clip]{line_6_xmm.eps}
 \caption[The O VIII Ly$\alpha$ region in {\it Chandra} spectra and {\it XMM-Newton} spectra]{The O VIII Ly$\alpha$ region in {\it Chandra} spectra (left) and {\it XMM-Newton} spectra (right) of CAL 87, and a fit, shown by the red solid line, using a Gaussian for the line and a power law for the continuum. The residuals are shown in the panel below each plot. The residuals show how the line profile in {\it Chandra} spectra can be interpreted as having an embedded absorption core at the centre of the line, rather than as double-peaked profiles.}
\label{fig:line 6}
\end{figure*}

\begin{figure}
\includegraphics[width=88mm]{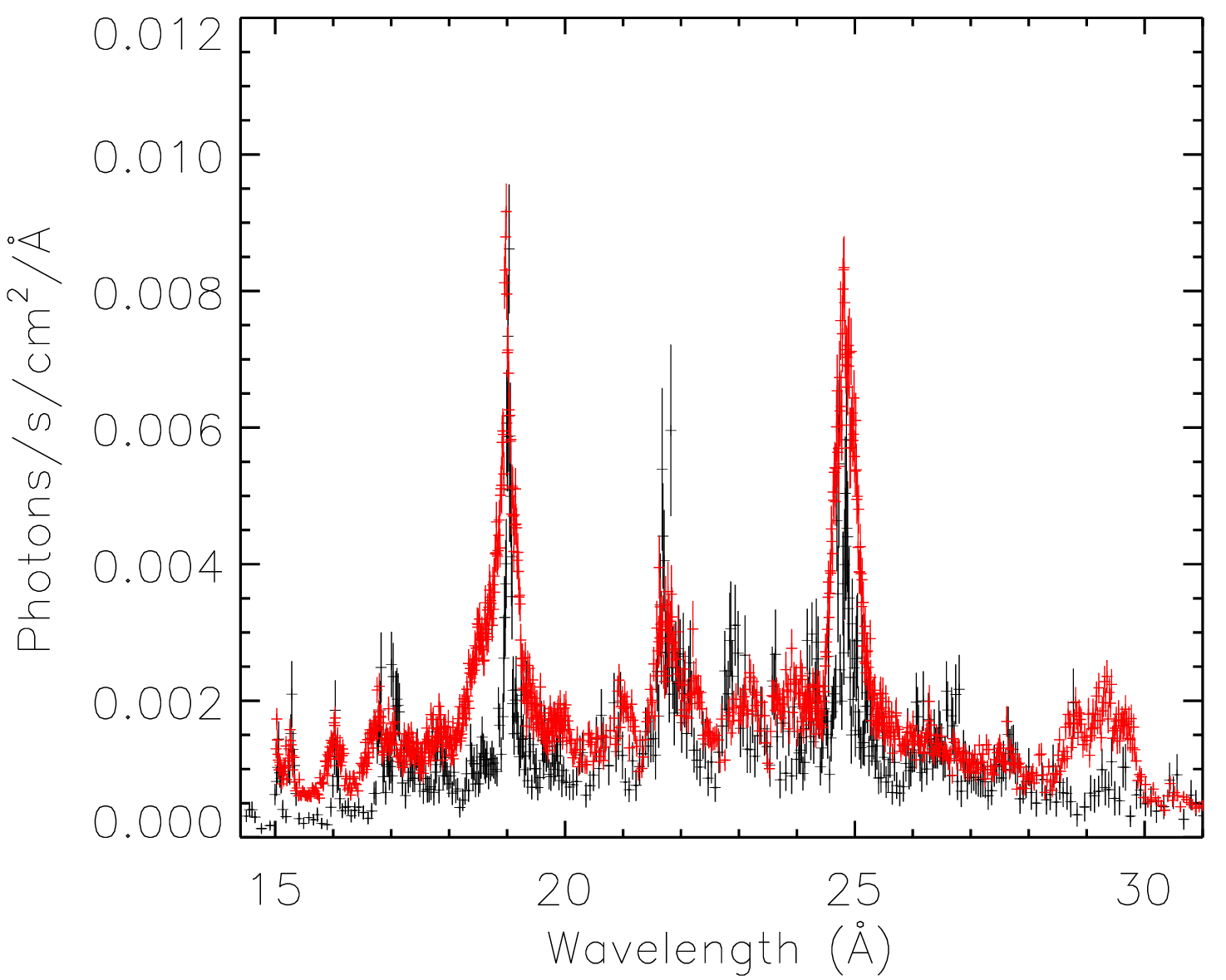}
 \caption[The spectrum of CAL 87 and U Sco]{The averaged {\it XMM-Newton} RGS spectrum of CAL 87 (black), multiplied
 by a factor of 7.7 to facilitate
 the comparison, and the RGS spectrum of U Sco in exposure no. 0561580301 done 35 
 days after the optical maximum of the 
 last outburst of this recurrent nova (red). 
}\label{fig:spectra cal87 usco}
\end{figure}

\section{Spectral analysis}
\label{sec:spectral}
Spectral fits of the high resolution X-ray spectra yield precious information on the nature of the system, but the high resolution X-ray spectra CAL 87 are complex. A fit was attempted in a short paper by \citet{2010AN....331..152E}, they found that in addition to the plasma in which the emission lines originate, the hot continuum corresponding to a WD is needed to explain the spectra, and they also noted that the WD continuum must be scattered by the Accretion Disc Corona (ADC) into the line of sight. \citet{2014ApJ...792...20R} focused on mapping the eclipse, and they found that the emission lines must come from an extended region.

Since the intensities of emission lines in the {\it Chandra} and {\it XMM-Newton} spectra do not show significant changes during/out of the eclipses \citep{2014ApJ...792...20R}, we fitted the spectra of the entire exposures. We also estimated the flux and determined the measured wavelengths of emission lines by using one or two Gaussian model (AGAUSS in XSPEC) to fit their profiles, with a simple power law model to describe the continuum around the short energy range of each line, and we determined the errors by using the error command in XSPEC. The identified emission lines and the
 flux, when measurable, are listed
 in Table~\ref{table:emission lines cal87} and labelled in Fig.~\ref{fig:bvapec}.
The intercombination and forbidden lines of the O VII triplet in the spectra of {\it XMM-Newton} data are too weak, so we could not estimate
 the errors in measuring the central wavelengths.
The N VI emission line in the spectra of {\it Chandra} data is too weak to
 calculate its flux.
We also compared the {\it Chandra} fluxes and spectra with those of {\it XMM-Newton}, and we found that there are some
 small differences, so we fitted the spectra of {\it Chandra} data and {\it XMM-Newton} data separately.

We obtained the best fits
 by combining the TMAP atmospheric model with galactic halo abundances developed by \citet{2000ASPC..199..337R} with a plasma in collisional ionization equilibrium (CIE, BVAPEC model in XSPEC, \citealt{2001ApJ...556L..91S}), or adding the TMAP model with galactic halo abundances to a plasma that is not in CIE (VNEI model in XSPEC, \citealt{2001ApJ...548..820B}). Both models can not explain all the lines and the unusual line ratios of Fe XVII. The best-fit parameters of these models are shown in Table~\ref{table:parameters bvapec cal87}, and the errors were estimated by using the error command in XSPEC. The parameters for the best fits obtained with the CIE and atmospheric components (Model 1) for both the {\it XMM-Newton} and {\it Chandra} spectra are listed in Table~\ref{table:parameters bvapec cal87}. Additionally, the parameters obtained with the NEI (non-equilibrium) model and atmospheric component (Model 2) are also included in Table~\ref{table:parameters bvapec cal87}. The fits to the spectra using an atmospheric model combined with a collisional ionization plasma model are shown in Fig.~\ref{fig:bvapec}, while the fits using an atmospheric model combined with a non-equilibrium ionization collisional plasma model are shown in Fig.~\ref{fig:vnei}. The table also gives the 90\% confidence level uncertainties. In the BVAPEC and VNEI models, the abundances of Fe, O, Ar and N are set as free parameters. However, since these abundances are not well constrained in the fits of best models, we have not included their values in Table~\ref{table:parameters bvapec cal87}. We attempted to add more components to the Model 1 and Model 2, such as a second component for the continuum and/or another region at different
 plasma temperature, but the fits to the spectra did not significantly improve. For instance, the fits for the {\it XMM-Newton} spectra yielded slightly smaller reduced $\chi^2$ values (1.96 for Model 1 and 2.01 for Model 2) when an additional atmospheric component was added, though the changes in the fits are hardly noticeable to the naked eye. The fit for the {\it XMM-Newton} spectra yielded a reduced $\chi^2$ value of 1.95 when an additional BVAPEC model was added to the Model 1, with only a marginal improvement in the 26 $-$ 27 \AA\ range. Similarly, the fit yielded a reduced $\chi^2$ value of 1.99 when an additional VNEI model was added to the Model 2, with only marginal improvement in the 26 $-$ 28 \AA\ range. In both cases, several other emission features remain unaccounted for.

As shown in Table~\ref{table:parameters bvapec cal87}, the unabsorbed X-ray flux of CAL 87 in the $0.2-1$ keV range 
with the CIE plasma and atmospheric model turns out to be 
20.0$^{+1.4}_{-1.5}$ $\times 10^{-13}$ erg cm$^{-2}$ s$^{-1}$ and 
21.5$^{+1.7}_{-1.9}$ $\times 10^{-13}$ erg cm$^{-2}$ s$^{-1}$ for
 the {\it Chandra} and {\it XMM-Newton}
 spectrum, respectively, and the emission lines contribute to at least 30\% of the X-ray flux. The absolute
 X-ray luminosity of CAL 87 in the $0.2-1$ keV range 
with the CIE plasma and atmospheric model turns out to be 
4.64$^{+0.28}_{-0.29}$ $\times 10^{36}$ erg s$^{-1}$ and 
4.54$^{+0.56}_{-0.53}$ $\times 10^{36}$ erg s$^{-1}$ for
 the {\it Chandra} and {\it XMM-Newton}
 spectrum, respectively. Assuming the non-equilibrium collisional plasma
 component and the atmospheric component,
 the X-ray luminosity turns out to be
 higher by several percent,
 5.46$^{+0.38}_{-0.37}$ $\times 10^{36}$ erg s$^{-1}$ and 
4.82$^{+0.40}_{-0.42}$ $\times 10^{36}$ erg s$^{-1}$ , for {\it Chandra}
 and {\it XMM-Newton} respectively. The absolute X-ray luminosity of CAL 87 in the $0.2-1$ keV range constitutes $\sim 95\%$ of the bolometric luminosity.

The atmospheric models developed by \citet{2000ASPC..199..337R} represent a good approximation to the atmosphere of a hydrogen burning WD, and including a grid of effective gravity values, with a log (g) grid ranges from 5 to 9 with 1 step increments. We do not have sufficient spectral leverage in our data to constrain the atmospheric continuum. In fact, because the level of the continuum flux is much lower than that of a hydrogen burning WD, the flux is not a limit that allows us to properly constrain simultaneously two parameters, log(g) and T$_{\rm eff}$. Moreover, because we do not detect absorption features, and have the overlap of the prominent emission lines, it is not possible to determine a suitable value of log(g) between values log(g)=6 to log(g)=9, because the WD luminosity and radius depend on the amount of N(H) that we assume. The hydrogen column density obtained in our fits is consistent with the value in the direction of CAL 87 that is 3.42 $\times$ 10$^{21}$ cm$^{-2}$ indicated by \citet{2016A&A...594A.116H}. We do obtain one important constraint by fitting the atmospheric model, that T$_{\rm eff}$ should be $\sim$ 800,000 K for any log(g). We calculated that, with log(g) as low as 6, we would have a very bloated radius with respect to a WD in hydrostatic equilibrium (7.5 $\times$ 10$^{9}$ cm to 10$^{10}$ cm). Model calculations show that at such high T$_{\rm eff}$ the radius has to be sufficiently small for the WD to remain in hydrostatic equilibrium, so with log(g) $\leq$ 8, the WD would actually be undergoing a nova outburst \citep[e.g.,][]{2012BaltA..21...76S}. The only stable configuration is obtained by assuming a radius close to that of a stable WD with T$_{\rm eff}$ $\simeq$ 800,000 K, which constrains log(g) to be close to 9. The putative WD would have has T$_{\rm eff}$ consistent with a WD mass of about 1.2 M \citep{2013ApJ...777..136W}. The unabsorbed X-ray luminosity inferred from our spectral ﬁts is very low compared to a hydrogen burning WD with effective temperature of 800,000 K, and a mass of $\sim$ 1.2 M$_\odot$, this implies that the continuum flux is attributed to a Thomson scattering corona. The amount of Thomson reflected flux of the WD would be about 5\% for log(g)=9. An eclipsed Thomson scattering corona around a WD was also suggested in nova U Sco \citep{2012ApJ...745...43N, 2013MNRAS.429.1342O}, where the amount of reprocessed radiation in different observations was 10\% to 20\%. The inference that the WD radiation was scattered was also suggested for nova HV Cet \citep{2012A&A...545A.116B} and Nova Monocerotis 2012 \citep{2013ApJ...768L..26P}.

\begin{table*}
 \centering
 \begin{minipage}{155mm}
  \caption[Parameters of emission lines identified]{Rest, measured wavelength, line shifts, and fluxes $\times$ $10^{-13}$ergs cm$^{-2}$ s$^{-1}$ for the emission lines identified and measured in the grating spectra of CAL 87. Fluxes are calculated with the cflux command in XSPEC. The errors are at a 90 percent confidence level.}
\label{table:emission lines cal87}
\begin{tabular}{@{}lcccc|ccc@{}}
\hline
Line               &  $\lambda_0$ (\AA)     & $\lambda_m$ (\AA) & $v_{\rm shift}$ (km\,s$^{-1}$)   & Flux &
$\lambda_m$ (\AA) & $v_{\rm shift}$ (km\,s$^{-1}$) & Flux \\
\hline
& & \multicolumn{3}{c}{2001-08-13/14}  &\multicolumn{3}{c}{2003-04-18/19}\\
& & \multicolumn{3}{c}{ {\it Chandra}\,ACIS+LETG}  &\multicolumn{3}{c}{ {\it XMM-Newton}\,RGS}\\
\hline
Fe XVII  & 15.0140 & 15.08$^{+0.06}_{-0.06}$                & 1317$^{+1180}_{-1189}$     & 0.20$^{+0.10}_{-0.10}$  & 15.05$^{+0.01}_{-0.01}$    &  719$^{+201}_{-198}$     & 0.23$^{+0.06}_{-0.05}$  \\
Fe XVII?  & 15.2610 & 15.32$^{+0.03}_{-0.04}$               & 1159$^{+589}_{-785}$    & 0.25$^{+0.09}_{-0.08}$  & 15.29$^{+0.01}_{-0.01}$     &  570$^{+198}_{-194}$     & 0.33$^{+0.07}_{-0.06}$ \\
O VIII Ly$\beta$   & 16.0055 &  16.05$^{+0.08}_{-0.07}$     &  834$^{+1498}_{-1311}$    & 0.18$^{+0.15}_{-0.13}$ & 16.04$^{+0.02}_{-0.01}$      &  646$^{+374}_{-187}$        & 0.22$^{+0.05}_{-0.05}$ \\
Fe XVII  & 16.7800 & 16.81$^{+0.05}_{-0.05}$                &  540$^{+890}_{-885}$     & 0.12$^{+0.08}_{-0.08}$ &  16.82$^{+0.01}_{-0.02}$    &  715$^{+179}_{-351}$     & 0.23$^{+0.07}_{-0.05}$  \\
Fe XVII  & 17.0510  & 17.11$^{+0.03}_{-0.03}$               &  1037$^{+525}_{-519}$     & 0.40$^{+0.12}_{-0.13}$  & 17.09$^{+0.01}_{-0.02}$    &  686$^{+178}_{-351}$     & 0.43$^{+0.09}_{-0.06}$ \\
O VIII Ly$\alpha$  & 18.9671   &  19.03$^{+0.03}_{-0.04}$   &  994$^{+472}_{-630}$   & 0.85$^{+0.12}_{-0.13}$  & 19.03$^{+0.01}_{-0.01}$      &  994$^{+154}_{-161}$         & 1.17$^{+0.08}_{-0.09}$   \\
 O VII r & 21.6015 & 21.75$^{+0.09}_{-0.08}$                &  2060$^{+1249}_{-1110}$     &0.54$^{+0.44}_{-0.42}$ &  21.69$^{+0.01}_{-0.01}$     &  1228$^{+141}_{-137}$         & 0.46$^{+0.12}_{-0.12}$ \\
 O VII i & 21.8010 &  $...$  & $...$     &  $...$                                                                     & 21.82                      &  261    & 0.14$^{+0.09}_{-0.09}$  \\
 O VII f & 22.0974 &  $...$   &  $...$    &  $...$                                                                    &     22.16               &  851          & 0.22$^{+0.11}_{-0.10}$ \\
N VI   & 22.8070 & 22.85$^{+0.05}_{-0.04}$              &  565$^{+651}_{-523}$   &    $...$   & 22.86$^{+0.02}_{-0.02}$                    &  697$^{+256}_{-260}$           & 0.26$^{+0.09}_{-0.07}$ \\
Ar XVI   & 23.5060 & 23.61$^{+0.08}_{-0.09}$                &  1326$^{+1018}_{-1143}$   &  0.11$^{+0.09}_{-0.09}$  &  23.61$^{+0.03}_{-0.04}$    &  1330$^{+386}_{-510}$          & 0.16$^{+0.07}_{-0.08}$ \\
N VII Ly$\alpha$   & 24.7792 & 24.84$^{+0.02}_{-0.05}$      &  736$^{+246}_{-602}$   & 0.48$^{+0.21}_{-0.21}$  & 24.86$^{+0.01}_{-0.01}$      &  978$^{+119}_{-124}$       & 0.48$^{+0.07}_{-0.07}$ \\

\hline
\end{tabular}
\end{minipage}
\end{table*}

\begin{table*}
 \centering
 \begin{minipage}{155mm}
\caption[Physical parameters of the fits]{Physical parameters of the fits for the spectra of CAL 87 shown in Fig.~\ref{fig:bvapec} (model atmosphere + collisional ionization plasma model BVAPEC in XSPEC, Model 1) and Fig.~\ref{fig:vnei} (model atmosphere + non-equilibrium ionization collisional plasma model VNEI in XSPEC, Model 2).
N(H) is the hydrogen column density. T$_{\rm WD}$ is the WD effective temperature in the atmospheric model. The temperature grid in the atmospheric model is interpolated in steps of 100,000 K, which reflects the realistic value of the uncertainty. F$_{\rm WD}$ is the WD atmospheric flux.
L$_{\rm WD}$ is the WD atmospheric luminosity.
T$_{\rm p}$ is the plasma temperature in either the BVAPEC model or the VNEI model.
The emission lines were also redshifted with redshift z around a few $\times$ 10$^{-3}$. Velocity$_{\rm p}$ is the broadening velocity of BVAPEC, and the parameter broadening velocity does not exist in the VNEI model. F$_{\rm p}$ is the collisional ionization plasma flux or the non-equilibrium ionization collisional plasma flux.
L$_{\rm p}$ is the collisional ionization plasma luminosity or the non-equilibrium ionization collisional plasma luminosity.
F$_{\rm tot}$ is the total flux.
L$_{\rm tot}$ is the total luminosity.
Absorbed and unabsorbed fluxes in the 0.2 $-$ 1.0 keV energy range are calculated with the cflux command in XSPEC.
The distance of 49.97 $\pm$ 1.11 kpc was used in the luminosity calculation \citep{2013Natur.495...76P}.
}
\label{table:parameters bvapec cal87}
  \begin{tabular}{@{}lcc|cc@{}}
  \hline
& \multicolumn{2}{c}{2001-08-13/14}  &\multicolumn{2}{c}{2003-04-18/19}\\
& \multicolumn{2}{c}{ {\it Chandra}\,ACIS+LETG}  &\multicolumn{2}{c}{ {\it XMM-Newton}\,RGS}\\
\hline
Parameter&   Model 1 & Model 2 & Model 1 & Model 2 \\
\hline
N(H) (10$^{21}$ cm$^{-2}$)   & 1.9$^{+0.3}_{-0.3}$ & 2.1$^{+0.3}_{-0.3}$  &   2.0$^{+0.4}_{-0.4}$  &  2.1$^{+0.4}_{-0.4}$  \\
T$_{\rm WD}$ (K)  & 796619 & 801179  & 796596    &  800440 \\
F$_{\rm WD}$ (abs.) (erg cm$^{-2}$ s$^{-1} \times 10^{-13}$)  & 16.0$^{+1.1}_{-1.1}$ & 16.7$^{+1.1}_{-1.1}$  &   16.0$^{+1.4}_{-1.4}$  & 16.3$^{+1.3}_{-1.2}$\\
F$_{\rm WD}$ (unabs.) (erg cm$^{-2}$ s$^{-1} \times 10^{-13}$)  & 112.9$^{+11.1}_{-10.8}$ & 129.5$^{+11.1}_{-12.0}$  & 119.4$^{+13.6}_{-13.8}$ & 126.9$^{+13.6}_{-14.8}$ \\
L$_{\rm WD}$ (unabs.) (erg s$^{-1} \times 10^{35}$)  & 33.7$^{+3.6}_{-3.5}$ & 38.7$^{+3.7}_{-4.0}$  &  35.7$^{+4.4}_{-4.4}$  & 37.9$^{+4.4}_{-4.7}$ \\
T$_{\rm p}$ (KeV)  & 0.24$^{+0.05}_{-0.06}$ & 0.22$^{+0.05}_{-0.05}$  & 0.23$^{+0.08}_{-0.07}$   & 0.25$^{+0.08}_{-0.07}$  \\
Redshift$_{\rm p}$(10$^{-3}$)   & 3.7$^{+0.1}_{-0.2}$ & 3.5$^{+0.1}_{-0.1}$  & 2.7$^{+0.1}_{-0.1}$   & 2.8$^{+0.1}_{-0.1}$  \\
Velocity$_{\rm p}$(km s$^{-1}$)  & 754$^{+118}_{-122}$ & $...$  & 514$^{+102}_{-94}$   & $...$  \\
F$_{\rm p}$ (abs.) (erg cm$^{-2}$ s$^{-1} \times 10^{-13}$)  & 4.5$^{+0.6}_{-0.6}$ & 3.5$^{+0.5}_{-0.6}$  & 5.5$^{+0.6}_{-0.6}$   &  5.0$^{+0.7}_{-0.7}$  \\
F$_{\rm p}$ (unabs.) (erg cm$^{-2}$ s$^{-1} \times 10^{-13}$)  & 44.9$^{+8.9}_{-9.7}$  &  53.5$^{+10.4}_{-12.6}$ & 32.0$^{+12.7}_{-13.3}$  & 33.5$^{+9.8}_{-10.1}$ \\
L$_{\rm p}$ (unabs.) (erg s$^{-1} \times 10^{35}$)  & 13.4$^{+2.8}_{-3.0}$ & 16.0$^{+3.2}_{-3.9}$   & 9.6$^{+3.8}_{-4.0}$  &  10.0$^{+2.9}_{-3.0}$ \\
F$_{\rm tot}$ (abs.) (erg cm$^{-2}$ s$^{-1} \times 10^{-13}$)   & 20.0$^{+1.4}_{-1.5}$ & 20.3$^{+1.5}_{-1.5}$  & 21.5$^{+1.7}_{-1.9}$   & 21.3$^{+1.8}_{-1.8}$ \\
F$_{\rm tot}$ (unabs.) (erg cm$^{-2}$ s$^{-1} \times 10^{-13}$)  &  155.4$^{+9.4}_{-9.9}$ & 182.9$^{+12.8}_{-12.5}$  & 151.9$^{+18.8}_{-17.8}$   &  161.5$^{+13.5}_{-14.0}$  \\
L$_{\rm tot}$ (unabs.) (erg s$^{-1} \times 10^{35}$)  & 46.4$^{+3.5}_{-3.6}$  &  54.6$^{+4.5}_{-4.4}$ &    45.4$^{+6.0}_{-5.7}$   &  48.2$^{+4.5}_{-4.7}$  \\
$\chi^2_{red}$  & 0.83  &  1.02 &   1.98     &  2.04 \\
\hline
\end{tabular}
\end{minipage}
\end{table*}

\section{Discussion}
\label{sec:discussion}
The emission lines are broadened, probably because the emission region is extended, and they are redshifted. The O VII He-like triplet can be used
 as a diagnostic to constrain the plasma.
In the {\it Chandra} spectrum we observe only the strong resonance line, and the forbidden and intercombination lines are absent in the {\it Chandra} spectrum,
 but the other lines of the triplet are measurable in
 the RGS spectrum, as shown in Fig.~\ref{fig:triplet}. This is due to the smaller
 collecting area of the {\it Chandra} LETG $+$ ACIS setup with respect to
 the RGS1 and RGS2 on {\it XMM-Newton}. ACIS and the LETG receive a lower photon count than the RGS1 and RGS2. 
 
Depending on the plasma temperature and specific element, the so-called G ratio indicates whether the plasma is in CIE, G = ${(f+i)}/r$, where $r$, $i$ and $f$ are the fluxes in the resonance, intercombination and forbidden lines of the He-like triplets, respectively, and another useful index of the plasma property is the R = $f/i$, which is the sensitive indicator of electron density. Generally, G $>$ 4 indicates a contribution of photoionization \citep{2000ApJ...544..581B, 2001A&A...376.1113P} as long as this diagnostic is used in a regime where the forbidden line is not sensitive to the density; that is, to extremely high densities. 
For the O VII He-like triplet, we measured G = 0.78 $\pm$ 0.35 and R = 1.57 $\pm$ 1.23. Even taking into account the errors in the measurement as we see in Fig.~\ref{fig:triplet}, the $r$ line is stronger than the other two of the O VII He-like triplet, so the G ratio value is much smaller
 than 4. These diagnostics point against a photoionized plasma,
 and possibly toward a plasma in CIE. The value of the R ratio of the O VII He-like triplet, corresponds to the electron density n$_{\rm e}\sim$4$\times$10$^{10}$ cm$^{-3}$ \citet{2000A&AS..143..495P}.

 However, the spectrum may be more complex and the plasma may
 not be in CIE. For the 
 {\it Chandra} spectrum, the non-equilibrium collisional plasma component 
 gives a slightly better fit than the CIE plasma,
 but the opposite is true for the {\it XMM-Newton} spectrum.
 There are astrophysical phenomena in which 
non-equilibrium ionization is indeed thought to occur: 
supernova remnants \citep{1982A&AS...48..305G, 1994ApJ...437..770M}, 
the outskirts of galaxy clusters, the inner
 part of galaxy clusters \citep{2008PASJ...60L..19A, 2010A&A...509A..29P}.

Another puzzle in the spectra of CAL 87 is due to an apparently
 very unusual line ratio in two Fe-L lines. The observed Fe XVII $I(15.01~{\rm \AA })/I(15.26~{\rm \AA })$ line ratio is 0.70$-$0.80 that is higher than the values in the theoretical calculations, in the experiments and in astrophysical sources, as shown in able~\ref{table:Fe XVII}. Fe-L lines are very common in the X-ray emission line spectra of novae
 and other SSSs. They
 are basically a complex assembly of $n\geq3$ to $n=2$ transitions of 
Fe ions in different ionization states, 
affected by a range of atomic processes such as collisional excitation, resonant excitation, radioactive recombination, dielectronic recombination, and inner-shell ionization.
 These lines are often very bright, and are frequently used as diagnostics of electron temperature (e.g., \citealt{1985ApJ...298..898S}), electron density (e.g., \citealt{1996ApJ...466..549P}), and chemical abundances \citep{2006A&A...459..353W, 2017A&A...607A..98D}.

 The rich science of Fe-L has motivated many studies, in particular 
for \ion{Fe}{XVII}. The L-shell line emission spectrum of \ion{Fe}{XVII} is observed over a large temperature range.
Two of the strongest, most distinct lines are the ($1s^22s^22p^5_{1/2}3d_{3/2})_{J=1}$ $\rightarrow$ ($1s^22s^22p^6)_{J=0}$ resonance and ($1s^22s^22p^5_{3/2}3d_{5/2})_{J=1}$ $\rightarrow$ ($1s^22s^22p^6)_{J=0}$ intercombination 
line at 15.014 and 15.261 \AA \, respectively (commonly labelled 3C and 3D).
As Fig.~\ref{fig:bvapec} shows, we tentatively identified two lines as
 such. These lines are measured 
 respectively at 15.32 \AA\ in the {\it Chandra} spectrum
 and 15.29 \AA\ in the {\it XMM-Newton} spectrum, and 
 at 15.08 \AA\ in {\it Chandra} and 15.05 \AA\ in {\it XMM-Newton}. Considering the uncertainties in wavelength or velocity, we further investigated the uniqueness of this identification. We noted that there are no other known emission lines in this wavelength region with sufficient intensity to match these two observed lines, but these lines could be due to a range of nearby ionization states. However, there remains a possibility that the line observed at 15.29 \AA\ and 15.32 \AA\ represent a new, as yet unidentified line. This topic will be revisited later in the text. In the literature there is some uncertainty
 in the ratio of the relative intensity of these
 two Fe XVII emission lines, and we summarize the published results 
for the line ratios in Table~\ref{table:Fe XVII}.
 In fact, the ratios of the lines can also completely change if the plasma density is the above a critical density that is a few $\times 10^{13}~\rm cm^{-3}$ or the photoexcitation temperature is above 55 kK \citep{2001ApJ...560..992M}. We do not, however, reach a clear conclusion that whether this high plasma density or high photoexcitation temperature regime described by \citet{2001ApJ...560..992M} encompasses the 0.7-0.8 values of Fe XVII $I(15.01~{\rm \AA })/I(15.26~{\rm \AA })$ line ratio we have observed.

 Despite
 the uncertainties, the 3C line in the theoretical calculations,
 in the experiments and in astrophysical sources has always been found to be stronger 
 than the 3D, unlike in our case: we measure a ratio 0.70$-$0.80. This was noticed also by \citet{2004RMxAC..20...18G}, who attributed this to recombination emission. There may be different explanations for this puzzle.
First of all, the
 contamination of the 3D ($\lambda_0=15.261$\,\AA) line by an 
Fe XVI satellite line can further reduce the value of 
 the ratio, especially in colder plasmas (below $\sim$ 3 MK) \citep{1997A&A...324..381P, 2001ApJ...557L..75B}. As shown in the Table~\ref{table:parameters bvapec cal87}, our spectral fitting results show that the plasma temperature is in the range of 0.22$-$0.25 keV (corresponding to 2.55$-$2.90 MK). Consequently, the unusually low ratio observed in CAL 87 may be due to an Fe XVI satellite line coinciding with the 3D line. \citet{2001ApJ...548..966B} also found that the attenuation of the ratio measured in Capella may be due to the blending of the Fe XVII ($\lambda_0=15.261$\,\AA) line with a strong Fe XVI ($\lambda_0=15.262$\,\AA \, measured at 15.264 $\AA$ in the spectra of Capella) line that can contribute to
about 10\% of the total intensity.

Another reason may have to do with unusually high density.
\citet{2001ApJ...560..992M} found that the Fe XVII $I(17.10~{\rm \AA })/I(17.05~{\rm \AA })$ line ratio (which we cannot measure for CAL 87),
 observed in the {\it Chandra} HETG spectrum of the intermediate polar (IP) EX Hya is $0.05\pm 0.04$, which is significantly smaller than $0.93\pm 0.11$ observed in the Sun. However, they found that the line ratios of all other lines
 [$I(15.01~{\rm \AA })/I(16.78~{\rm \AA })$, $I(15.26~{\rm \AA })/I(16.78~{\rm \AA })$, $I(17.05~{\rm \AA })/I(16.78~{\rm \AA })$ and $I(15.01~{\rm \AA })/I(15.26~{\rm \AA })$] (thus including the 3C and 3D line ratio) in EX~Hya and the Sun are consistent with each other, as shown 
 also in Table~\ref{table:Fe XVII}. These
authors found that the significantly small value of the $I(17.10~{\rm \AA })/I(17.05~{\rm \AA })$ line ratio observed in EX~Hya can be explained if the plasma density $n\geq 2\times 10^{14}~\rm cm^{-3}$, that is orders of magnitude greater than that observed in the Sun or other late-type stars. 
Although there are
 no calculations for the relative intensity of the 3C and 3D Fe XVII lines,
 the fact that there is a critical density above which
 the ratio of the lines may completely change is
 well established \citep{2004ApJ...610..616B}.

We note that the two lines are also measured
with the same unusual ratio we found in Table~\ref{table:emission lines cal87} in the grating spectra of another SSS, MR Vel \citep{2002A&A...385..511B}.
If our line identification is correct, given
that even a density $n\geq 2\times 10^{14}~\rm cm^{-3}$ did not skew
 the ratio of these lines in EX Hya, this unusual measurement may be due to the Fe XVII line
 at 15.261 \AA \ being blended with a stronger than usual Fe XVI line at
 15.262 \AA. However, we also do not rule
 out that we did not measure the 3D line and that this
 is instead a new, still unidentified line. We think that new calculations and
 experiments for the L-shell line emission spectrum of Fe XVII may still be
 needed. Furthermore, also the ratio of Fe XVII at 16.78 \AA \ and the lines at 17.05 \AA \ is unusually small. This was noticed also by \citep{2004RMxAC..20...18G}, who attributed this to recombination emission. All the iron lines we could identify have very unusual ratios and are not consistent with a CIE plasma.

There is another interesting finding: a possible
narrow absorption line inside the broadened emission lines of O VIII Ly$\alpha$ and N VII Ly$\alpha$ in the {\it Chandra} spectra but not in the {\it XMM-Newton} spectra. Fig.~\ref{fig:line 6} shows this finding in the emission line of O VIII Ly$\alpha$ as an example illustrating how the line profile can be interpreted as having an embedded absorption core at the centre of the line. As shown in Fig.~\ref{fig:line 6}, we can clearly see the absorption in the residual of O VIII Ly$\alpha$ region in {\it Chandra} spectra, and there is no this feature in the residual of O VIII Ly$\alpha$ region in {\it XMM-Newton} spectra. The narrow absorption lines inside the broadened emission lines of O VIII Ly$\alpha$ and N VII Ly$\alpha$ in the {\it Chandra} spectra are present during/out of the eclipses. The longer wavelength part of the emission line O VIII Ly$\alpha$ (in the right side of the absorption core which is at $\sim$ 18.96 \AA\ in Fig.~\ref{fig:line 6}) in the {\it Chandra} spectra has a clearly double peaked profile in the 19.0$-$19.1 \AA\ range, this double peaked profile was also noticed by \citet{2004RMxAC..20...18G}. We refer to this double peaked profile in the rest text. For the narrow absorption line inside the broadened emission line, the dip reached down to $<$ 20\% of the flux of the peak on the left-hand side, and an apparent inverted Gaussian profile can be observed in the same region as the narrow absorption line in the residuals when fitting the spectra with a Gaussian for the line and a power law for the continuum, as shown in Fig.~\ref{fig:line 6}. In contrast, for the double peaked profile, the dip dropped to $>$ 80\% of the flux of the left-hand peak, and no apparent inverted Gaussian profile is present in the same region as the double peaked profile in the residuals when using the same fitting method. As we know, the spectral resolution of the RGS is less than ACIS+LETG, but double-peaked profiles were detected in the broadened emission lines of O VIII Ly$\alpha$ of {\it Chandra} spectra and {\it XMM-Newton} spectra \citep{2004RMxAC..20...18G, 2014ApJ...792...20R}, and these double-peaked profiles are more fine structures than the narrow absorption line inside the broadened emission line of O VIII Ly$\alpha$, therefore, lower spectral resolution of the RGS should be not the reason that there are no absorption lines in the broadened emission lines of the {\it XMM-Newton} spectra. These absorption lines are at their rest wavelengths (unshifted). We suggest that there was some rest and transient material that is outside of the emitting region of the emission lines.

The emission lines in {\it Chandra} spectra are clearly broad and redshifted with velocities in the range from 540 to 2060 km s$^{-1}$ (see Table~\ref{table:emission lines cal87}). 
The emission lines in {\it XMM-Newton} spectra of 2003 are also broad and redshifted 
with velocities in the range from 261 to 1330 km s$^{-1}$, as shown in Table~\ref{table:emission lines cal87}, mostly
 less redshifted than
the corresponding emission lines in the {\it Chandra} spectra of 2001. As shown in the Table~\ref{table:parameters bvapec cal87}, 
 the resulting average redshift for the emission lines in the {\it Chandra} spectrum 
is z = 3.70$^{+0.13}_{-0.16}$ $\times$ $10^{-3}$ (corresponding to 1109$^{+39}_{-48}$ km/s)
 in the first model and 
 z = 3.51$^{+0.11}_{-0.12}$ $\times$ $10^{-3}$ (corresponding to 1052$^{+33}_{-39}$ km/s)
 in the second.
 The average redshift resulting from
 the fit to the {\it XMM-Newton} spectra is z = 2.70$^{+0.12}_{-0.11}$ $\times$ $10^{-3}$ (corresponding to 809$^{+36}_{-33}$ km/s) and z = 2.75$^{+0.10}_{-0.12}$ $\times$ $10^{-3}$ (corresponding to 824$^{+30}_{-36}$ km/s) for each of the two models
, respectively. They are also broadened by about 750 km s$^{-1}$ in the LETG spectrum and by about 500 km s$^{-1}$ in the RGS one. The stronger emission lines in CAL 87 are observed at wavelengths where the continuum is more intense, indicating photoexcitations as part of resonant line scattering \citep{2013A&A...559A..50N}.

The significant red shift has been interpreted by \citet{2004RMxAC..20...18G} as implying
 that the emission lines emanate in a wind.
 If we can observe all parts of a spherically symmetric wind, we should observe broadened lines centered on the systemic velocity. 
If the outflow is due to an almost point source, the blue shifted portion 
 of the wind coming towards the line of sight is negligible compared to
 the redshifted outflow. However, \citet{2004RMxAC..20...18G} noticed that
 the wind may also come from an extended accretion disk. They also found that the O VIII line at 18.97 \AA \ in the {\it Chandra} spectrum has a clearly double peaked profile, but the red peak disappears during the eclipse.
 This can be interpreted and modelled as an outflow in a bi-directional cone
 of an accretion disk corona, with an opening angle of about 120$^o$ \citep{2004RMxAC..20...18G}.

The phase duration of the eclipse is 0.3 (see Fig.~\ref{fig:lc_cal87}), implying that the radius of the eclipsed source is much bigger than the radius of the WD. The eclipsed source should have about the
 same size
 as the occulter \citep{2004RMxAC..20...18G} and assuming
 that the secondary star is filling its Roche lobe, and the radius of the eclipsed source is about 1.4 R$_\odot$. 
In fact, since the orbital period is 10.6 hours, the effective temperature from our best spectra fitting implies that the mass of the WD is around 1.2 M$_\odot$.
If the mass of the secondary is around 0.4 M$_\odot$ \citep{1998ApJ...502..408H}, using Kepler's law the binary separation is about 2.9 R$_\odot$ and 
the radius of the Roche lobe is calculated following
\citet{1983ApJ...268..368E}. Thus, the eclipsed source is way too extended
to be a compact object. 
Since the flux of the emission lines does not decrease during the eclipse, they
 may originate in this extended region that is outside of the eclipsed region ($\sim$ 1.4 R$_\odot$). 

A similar result was obtained
 for the size of the emission region ($\sim$ 1.5 R$_\odot$) by \citet{2004RMxAC..20...18G}, even if these authors assumed
 that the mass of the WD is only 0.75 M$_\odot$ and that
 the secondary is more massive. The continuum component in the CAL 87 spectra is produced by the WD's photosphere \citet{2013A&A...559A..50N}. The fact that the continuum component is not totally eclipsed \citep{2014ApJ...792...20R} implies that
 also part of the continuum flux originates outside the eclipsed source. 
Therefore, like in the
recurrent nova U Sco, an extended corona around the WD, 
 most likely due to Thomson scattering, is needed to explain the X-ray spectra of CAL 87. Thomson scattering scatters the emission in a large volume without changing the spectrum. \citet{2013A&A...559A..50N} also suggest that the emission lines originate from the reprocessed emission far from the WD, in the wind.

The TMAP component in the spectral fit indicates that the effective temperature of CAL 87 is around 800,000 K. 
However, the effective temperature of RS Ophiuchi is also around 800,000 K, but
the X-ray luminosity of CAL 87
is $\sim 5 \times10^{36}$ ergs s$^{-1}$ in the $0.2-1$ keV range, which is a factor of 18 smaller than the X-ray luminosity in the same range measured in RS Ophiuchi \citep{2008ApJ...673.1067N}. This fact further suggests that
 we do not observe the WD directly in CAL 87
 and most of the emission is blocked by the accretion disk.
The reprocessing factor of the WD radiation should be in the order of 10\%. 
 This is consistent with the very high inclination angle \citep{1997A&A...321..245M, 2007A&A...472L..21O, 2014ApJ...792...20R}.

The spectra of CAL 87 are different from the spectra of other SSS such as CAL 83, because of the lack of strong absorption lines. 
The flux in the observed energy range ($0.2-1$ keV) is very low for an
 SSS in the LMC \citep{2006csxs.book..461K}, while the fluxes of the emission lines are unusually high for a system that does not have any
 trace of a recent nova shell at optical wavelengths \citep{2013A&A...559A..50N}.
 However, as shown in Fig.~\ref{fig:spectra cal87 usco}, we found that the spectra of CAL 87 are similar to the spectrum of
 one particular nova, U Sco, 35 days after its last nova outburst, and this similarity was also found by \citet{2013A&A...559A..50N}. In U Sco, the SSS is only observed as a Thomson scattering corona, and the emission lines are attributed to the nova ejecta. It is interesting to notice that in both CAL 87 and U Sco, 
the continuum has been attributed to a Thomson scattering corona, and
 U Sco is the only SSS-nova whose spectrum has clear similarities with
 the X-ray spectrum of CAL 87.
Assuming that the compact object in CAL 87 is a WD, 
the inclination obtained by \citet{1997A&A...321..245M, 2007A&A...472L..21O,
 2014ApJ...792...20R} is in the 78$-$$82^{\circ}$ range.
Like CAL 87, U Sco shows an X-ray eclipse, and
 has an inclination angle $82.7^{\circ}$$ \pm $$2.9^{\circ}$ \citep{2001MNRAS.327.1323T}.
 The effective temperature ($\sim$ 900,000 K) of U Sco is consistent with a 
WD mass of at least 1.3 M$_\odot$ and the reprocessing factor for 
the WD radiation
 must have been of the order of 10\% \citep{2013MNRAS.429.1342O}. Our spectral fitting 
 indicates similar values for CAL 87,
 and also very similar absolute X-ray luminosity for both.

The effective temperature ($\sim$ 800,000 K) of CAL 87 is consistent with a WD mass of about 1.2 M$_\odot$ \citep[according to][]{2013ApJ...777..136W}. 
 We suggest that CAL 87 and U Sco are likely the same type of binary
 (high mass hydrogen burning WD
 viewed in a high inclination binary), with different mass accretion rates.
 It is reasonable to suggest that the mass accretion rate is lower in U Sco and higher in CAL 87, that
 does not undergo nova outbursts, while
 U Sco is a recurrent nova. There is no correlation between the mass accretion rate and the orbital period for novae \citep{2019A&A...622A.186S}, we therefore cannot obtain the relative accretion rates of these two system based on their relative orbital periods. Since CAL 87 may be an accreting WD without nova explosion, it is a type Ia progenitor candidate because the mass may increase one day to become close to the Chandrasekhar value.

\begin{table*}
\begin{minipage}{155mm}
\caption[The ratio of the relative intensity of Fe XVII emission lines]{The ratio of the relative intensity of Fe XVII emission lines at 15.014 and 15.261 Angstrom ($I(15.01~{\rm \AA })/I(15.26~{\rm \AA })$, also labeled as $I_{3C}/I_{3D}$).}
\label{table:Fe XVII}
\begin{center}
\scalebox{1.0}{
\begin{tabular}{lcc}\hline\hline \noalign{\smallskip}
Source &   Ratio [$I_{3C}/I_{3D}$] & References   \\ 
 
\hline

\multicolumn{3}{l}{Theory:}\\

\citealt{1992ADNDT..52....1B}   & 4.28   &  \\
\citealt{1994ADNDT..58....1C}   & 4.57   &   \\
\citealt{1997ApJS..108..389M}   & 3.99   &   \\
\citealt{2001PhRvA..64a2507S}   & 3.43   &   \\
\citealt{2007PhRvA..76f2708C}   & 3.54   &   \\
\citealt{2009arXiv0905.0519G}   & 3.50   &   \\
\citealt{2012Natur.492..225B}   & 3.49   &   \\
\citealt{2014PhRvL.113n3001O}   & 3.56   &  \\
\citealt{2015PhRvA..91a2502S}   & 3.44   &  \\
\citealt{2017PhRvL.118p3002M}   & 2.82   &  \\
\citealt{2019NatSR...9.7463W}   & 3.567   &  \\

\hline

\multicolumn{3}{l}{Laboratory:}\\

LCLS$^a$   & $2.61 \pm 0.23$   &   \citealt{2012Natur.492..225B}  \\
PLT$^b$   & $2.48 \pm 0.40$   &   \citealt{2001PhRvA..64c2705B}  \\
          & $2.13 \pm 0.37$   &   \citealt{2004ApJ...610..616B}  \\
EBIT$^c$  & $3.04 \pm 0.12$   &   \citealt{1998ApJ...502.1015B}  \\
         & $1.90 \pm 0.11$$^d$   &   \citealt{2001ApJ...557L..75B}  \\
         & $2.63 \pm 0.15$$^e$   &   \citealt{2001ApJ...557L..75B}  \\
         & $2.77 \pm 0.3$   &   \citealt{2006PhRvL..96y3201B}  \\
PolarX-EBIT$^f$ & 3.09 $\pm$ 0.08$_\mathrm{stat.}$ $\pm$ 0.06$_\mathrm{sys.}$   &   \citealt{2020PhRvL.124v5001K}  \\
AtomDB$^g$  & 3.22   &     \\
\hline

\multicolumn{3}{l}{Astrophysical sources:}\\

Sun          & $2.02\pm 0.28$   &   \citealt{1999ApJ...510.1064S}  \\
NGC 4635      & $2.30 \pm 0.18$   &   \citealt{2002ApJ...579..600X}  \\
Capella       & $2.42 \pm 0.22$   &   \citealt{2001ApJ...548..966B}  \\
              & $2.85 \pm 0.14$   &  \citealt{Mewe2001}  \\
EX Hya          & $2.46\pm 0.42$   &   \citealt{2001ApJ...560..992M}  \\
CAL 87          & 0.70$-$0.80   &   This work  \\
\hline  \noalign{\smallskip}
\end{tabular}
}
\end{center}
Notes:\hspace{0.1cm} $^a $: Linac Coherent Light Source (LCLS) \citep{2010NaPho...4..641E} free-electron X-ray laser. 
$^b $: Princeton Large Torus (PLT) tokamak. $^c $: Livermore electron beam ion trap (EBIT). $^d $: Injection condition is in the high-pressure case. $^e $: Injection condition is in the low-pressure case. $^f $: PolarX-electron beam ion trap (PolarX-EBIT) \citep{2018RScI...89f3109M}. $^g $: AtomDB database version 3.0.9 \citep{2001ApJ...556L..91S, 2012ApJ...756..128F, 2020Atoms...8...49F}. \\

\end{minipage}
\end{table*}

\section{Conclusions}
\label{sec:conclusions}
We analysed the puzzling high resolution X-ray spectra of CAL 87 by using the archive X-ray data observed with the {\it Chandra} and {\it XMM-Newton}. Our analysis reached two main results that were previously concluded, and these two results are summarized as follows:

$\bullet$ The emission lines originate from an extended region, that was previously proposed by \citet{2014ApJ...792...20R}.

$\bullet$ The continuum appears to be due to a Thomson scattered corona reflecting a small
 fraction of the X-ray flux of the WD, and Thomson scattering was previously proposed by \citet{2001ApJ...550.1007E, 2013A&A...559A..50N} to explain the X-ray spectra of CAL 87.

Our analysis reached several new main results that extend what was done previously for CAL 87. The new main results resulting from our analysis can be summarized as follows:

$\bullet$ There may be a new, unidentified emission line around 15.29 Angstrom.

$\bullet$ Two narrow absorption cores were found in the broadened emission lines in the spectra of August, 2001 but not in the spectra of April 2003.
This implies that there may be some transient material at
 rest wavelength that is outside of the emitting region of the emission lines. 

$\bullet$ The collisional ionization plasma in CAL 87 may not be in equilibrium.

$\bullet$ We used physical models to fit the grating spectra of CAL 87, and we found that the best model includes
a WD atmosphere, with surface gravity $\sim$ $10^{9}$ cm s$^{-1}$ , effective temperature $\sim$ 800,000 K and mass $\sim$ 1.2 M$_\odot$. Thus,
CAL 87 contains a massive WD and, if
 it has high mass accretion rate that prevents nova
 outbursts, it may be a type Ia supernova progenitor.

\section*{Acknowledgements}
We express our deep gratitude to the anonymous referee for her or his constructive comments and suggestions, which helped us to improve the scientific content of this article. This research has made use of data obtained with the gratings on board {\it Chandra} and {\it XMM-Newton}. Songpeng Pei thanks Marina Orio, Ehud Behar, Uria Peretz and Jan-Uwe Ness for many useful conversations. This work was funded by the High-level Talents Research Start-up Fund Project of Liupanshui Normal University (LPSSYKYJJ202208), Science Research Project of University (Youth Project) of the Department of Education of Guizhou Province (QJJ[2022]348), the Science and Technology Foundation of Guizhou Province (QKHJC-ZK[2023]442), the Discipline-Team of Liupanshui Normal University (LPSSY2023XKTD11), and the Research Foundation of Qiannan Normal University for Nationalities (No.QNSY2019RC02).

\bibliographystyle{aa}
\bibliography{CAL87}

\end{document}